\documentclass{article} 
\usepackage{nips15submit_e,times}
\usepackage{hyperref}
\usepackage{url, mdwlist, wrapfig, subcaption, graphicx, varwidth}
\usepackage{footnote}
\makesavenoteenv{tabular}
\makesavenoteenv{table}
\usepackage{listings}
\usepackage{array}
\newcolumntype{L}[1]{>{\raggedright\let\newline\\\arraybackslash\hspace{0pt}}m{#1}}
\newcolumntype{C}[1]{>{\centering\let\newline\\\arraybackslash\hspace{0pt}}m{#1}}
\newcommand{\hs}[1]{\hspace{#1ex}}

\lstdefinelanguage{Julia} {morekeywords={using}}
\title{MXNet: A Flexible and Efficient Machine Learning Library for
  Heterogeneous Distributed Systems
}

\author{
Tianqi Chen,\hs{1} Mu Li\thanks{Corresponding author (muli@cs.cmu.edu)},\hs{1} Yutian Li,\hs{1}
Min Lin,\hs{1} Naiyan Wang,\hs{1} Minjie Wang,\\
\hs{-5}U.~Washington\hs{1.5} CMU\hs{3} Stanford\hs{4} NUS\hs{6} TuSimple\hs{6} NYU \\
\bf Tianjun Xiao,\hs{1} Bing Xu,\hs{1} Chiyuan Zhang,\hs{1} Zheng Zhang\\
\hs{3}Microsoft\hs{2.5} U.~Alberta\hs{6} MIT\hs{7} NYU Shanghai
}

\nipsfinalcopy 

\begin{document}

\maketitle

\vspace{-4ex}
\begin{abstract}
  MXNet is a multi-language machine learning (ML) library to ease the
  development of ML algorithms, especially for deep neural networks.
  Embedded in the host language, it blends declarative symbolic expression with
  imperative tensor computation. It offers auto differentiation to derive
  gradients.
  MXNet is computation and memory efficient and runs on
  various heterogeneous systems, ranging from mobile devices to distributed GPU
  clusters.

  This paper describes both the API design and the system
  implementation of MXNet, and explains how embedding of both symbolic expression
  and tensor operation is handled in a unified fashion.
  Our preliminary experiments reveal promising results on
  large scale deep neural network applications using multiple GPU machines.
\end{abstract}

\section{Introduction}

The scale and complexity of machine learning (ML) algorithms are becoming
increasingly large. Almost all recent ImageNet challenge~\cite{RusDenSuKretal15}
winners employ neural networks with very deep layers, requiring billions of
floating-point operations to process one single sample. The rise of structural
and computational complexity poses interesting challenges to ML system design
and implementation.

Most ML systems embed a domain-specific language (DSL) into a host language
(e.g. Python, Lua, C++). Possible programming paradigms range from {\em imperative},
where the user specifies exactly ``how'' computation needs to be
performed, and {\em declarative}, where the user specification focuses on
``what'' to be done. Examples of imperative programming include numpy and
Matlab, whereas packages such as Caffe, CXXNet program over layer definition
which abstracts away and hide the inner-working of actual implementation. The dividing line between the two can be muddy at times. Frameworks
such as Theano and the more recent Tensorflow can also be viewed as a mixture of both, they declare a computational graph, yet the computation within the graph is imperatively specified.

Related to the issue of programming paradigms is how the computation is carried out.
Execution can be {\em concrete}, where the result is returned right away on the same thread,
or {\em asynchronize} or {\em delayed}, where the statements are gathered and transformed into a
dataflow graph as an intermediate representation first, before
released to available devices. These two execution models have different
implications on how inherent parallelisms are discovered.
Concrete execution is restrictive (e.g. parallelized matrix multiplication), whereas asynchronize/delayed execution additionally identified all
parallelism within the scope of an instance of dataflow graph automatically.

The combination of the programming paradigm and execution model yields a large
design space, some of which are more interesting (and valid) than others. In
fact, our team has collectively explored a number of them, as does the rest of
the community. For example, Minerva~\cite{WanXiaLiZhaetal14} combines imperative
programming with asynchronize execution. While Theano takes an declarative approach,
 enabling more global graph-aware optimization.
Similar discipline was adopted in Purine2~\cite{LinLiLuoYan14}.
Instead, CXXNet adopts declarative programming (over tensor abstraction) and concrete
execution, similar to Caffe~\cite{JiaSheDonKaretal14}. Table~\ref{tab:dsl} gives
more examples.

Our combined new effort resulted in {\em MXNet} (or ``mix-net''), intending to
blend advantages of different approaches. Declarative programming offers clear boundary on the global computation graph, discovering more optimization opportunity,
whereas imperative programs offers more flexibility.
In the context of deep learning, declarative programming is useful in specifying the computation structure in neural network configurations,
while imperative programming are more natural for  parameter updates and interactive debugging.
We also took the effort to embed into multiple host languages, including C++, Python, R, Go and Julia.

Despite the support of multiple languages and combination of different programming paradigm,
we are able to fuse the execution to the same backend engine.
The engine tracks data dependencies across computation graphs and imperative operations, and schedules them efficiently jointly.
We aggressively reduce memory footprint, performing in-place
update and memory space reuse whenever possible. Finally, we designed a compact communication API so
that a MXNet program runs on multiple machines with little change.

Comparing to other open-source ML systems, MXNet provides a superset programming
interface to Torch7~\cite{ColKavFar11}, Theano~\cite{BasLamPasBeretal12}, Chainer~\cite{Chainer} and
Caffe~\cite{JiaSheDonKaretal14}, and supports more systems such as GPU clusters.
Besides supporting the optimization for declarative programs as TensorFlow~\cite{AbaAgaBarBreetal15} do, MXNet additionally embed imperative tensor operations to provide more flexibility.
MXNet is lightweight, e.g.~the prediction
codes fit into a single 50K lines C++ source file with no other dependency, and
has more languages supports.
More detailed comparisons are shown in Table~\ref{tab:comp}.

\begin{wrapfigure}[5]{r}{.3\textwidth}
  \centering
  \vspace{-70px}
  \includegraphics[width=\linewidth]{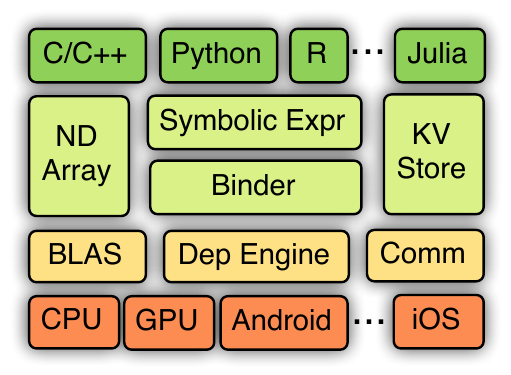}
  \vspace{-18px}
  \caption{MXNet Overview}
  \label{fig:arch}
\end{wrapfigure}

\begin{table}[t!]
    \centering
    \begin{tabular}{|L{1.3cm}|L{5.1cm}|L{6.4cm}|}
        \hline
        & Imperative Program & Declarative Program \\
        \hline
        Execute $a=b+1$ &
        Eagerly compute and store the results on $a$ as the same type with $b$. &
        Return a computation graph; bind data to $b$ and do the computation later. \\
        \hline
        Advantages &
        Conceptually straightforward, and often works seamless with the host
        language's build-in data structures, functions, debugger, and third-party libraries.
        &
        Obtain the whole computation graph before execution, beneficial for optimizing
        the performance and memory utilization.
        Also convenient to implement functions such as load, save, and visualization.
        \\
        \hline
    \end{tabular}
    \caption{Compare the imperative and declarative for domain specific languages. }
    \label{tab:dsl}
\end{table}

\begin{table}[t!]
  \centering
\setlength{\tabcolsep}{4pt}
  \begin{tabular}{|l|l|l|l|c|c|c|c|}
    \hline
     System & Core & Binding & Devices & Distri- & Imperative & Declarative  \\
     & Lang & Langs & (beyond CPU) & buted & Program & Program \\
     \hline
     Caffe~\cite{JiaSheDonKaretal14} & C++ & Python/Matlab & GPU &
     $\times$ & $\times$ & $\surd$ \\
     Torch7~\cite{ColKavFar11} &  Lua & - & GPU/FPGA & $\times$ & $\surd$ & $\times$ \\
     Theano~\cite{BasLamPasBeretal12} & Python & - & GPU & $\times$ & $\times$ & $\surd$ \\
     TensorFlow~\cite{AbaAgaBarBreetal15} & C++ & Python & GPU/Mobile &
     $\surd$
     & $\times$ & $\surd$ \\
     MXNet & C++ & Python/R/Julia/Go & GPU/Mobile & $\surd$ & $\surd$ & $\surd$ \\
     \hline
  \end{tabular}
  \caption{Compare to other popular open-source ML libraries}
  \label{tab:comp}
\end{table}

\section{Programming Interface}
\label{sec:progr-interf}

%
%

\subsection{\texttt{Symbol:} Declarative Symbolic Expressions}

MXNet uses multi-output symbolic expressions, \texttt{Symbol}, declare the
computation graph.
Symbols are composited by operators, such as simple matrix operations
(e.g. ``+''), or a complex neural network layer (e.g. convolution layer).
 An operator can take several input
variables, produce more than one output variables, and have internal state
variables. A variable can be either free, which we can bind with value later, or
an output of another symbol.
Figure~\ref{fig:julia} shows the construction of a multi-layer
perception symbol by chaining a variable , which presents the input data, and
several layer operators.

To evaluate a symbol we need to bind the free variables with data and declare
the required outputs. Beside evaluation (``forward''), a symbol supports auto
symbolic differentiation (``backward'').  Other functions, such as load, save,
memory estimation, and visualization, are also provided for symbols.

\begin{figure}[t!]
  \centering
  \begin{minipage}{.56\textwidth}
    \centering
    \lstset{language=Julia}
    \begin{lstlisting}[frame=single]
using MXNet
mlp = @mx.chain mx.Variable(:data) =>
  mx.FullyConnected(num_hidden=64) =>
  mx.Activation(act_type=:relu)    =>
  mx.FullyConnected(num_hidden=10) =>
  mx.Softmax()
    \end{lstlisting}
    \vspace{-10px}
    \caption{Symbol expression construction in Julia.}
    \label{fig:julia}
  \end{minipage}
  \begin{minipage}{.4\textwidth}
    \centering
    \lstset{language=Python}
    \begin{lstlisting}[frame=single]
>>> import mxnet as mx
>>> a = mx.nd.ones((2, 3),
... mx.gpu())
>>> print (a * 2).asnumpy()
[[ 2.  2.  2.]
 [ 2.  2.  2.]]
    \end{lstlisting}
    \vspace{-10px}
    \caption{NDArray interface in Python}
    \label{fig:python}
  \end{minipage}\hfill%
\end{figure}
\subsection{\texttt{NDArray:} Imperative Tensor Computation}

MXNet offers \texttt{NDArray} with imperative tensor computation to fill the gap between
the declarative symbolic expression and the host language.  Figure~\ref{fig:python} shows an
example which does matrix-constant multiplication on GPU and then prints
the results by \texttt{numpy.ndarray}.

\texttt{NDArray} abstraction works seamlessly with the executions declared by \texttt{Symbol}, we can mix
the imperative tensor computation of the former with the latter.
For example, given a symbolic neural network and the weight updating function,
e.g.~$w = w - \eta g$. Then we can implement the gradient descent by

\begin{figure}[th!]
\vspace{-6px}
\centering
\begin{varwidth}{\linewidth}
\begin{verbatim}
while(1) { net.foward_backward(); net.w -= eta * net.g };
\end{verbatim}
\end{varwidth}
\vspace{-6px}
\end{figure}

The above is as efficient as the implementation using a single but often much more complex symbolic expression.
The reason is that MXNet uses lazy  evaluation of \texttt{NDArray}  and the backend
engine can correctly resolve the data dependency between the two.

\subsection{\texttt{KVStore:} Data Synchronization Over Devices}

The \texttt{KVStore} is a distributed key-value store for data synchronization
over multiple devices.
%
It supports two primitives: {\em push} a key-value pair from a device to the store, and {\em pull} the
value on a key from the store. In addition, a user-defined
updater can specify how to merge the pushed value.
Finally, model divergence is controlled via consistency model~\cite{LiAndParSmoetal14}. Currently, we support the sequential and eventual consistency.

The following example implements the distributed gradient descent by data parallelization.

\begin{figure}[th!]
\vspace{-6px}
\centering
\begin{varwidth}{\linewidth}
\begin{verbatim}
while(1){ kv.pull(net.w); net.foward_backward(); kv.push(net.g); }
\end{verbatim}
\end{varwidth}
\end{figure}
\vspace{-6px}

where the weight updating function is registered to the \texttt{KVStore}, and
each worker repeatedly pulls the newest weight from the store and then pushes
out the locally computed gradient.

The above mixed implementation has the same performance comparing to a single
declarative program, because the actual data push and pull are executed by lazy
evaluation, which are scheduled by the backend engine just like others.

\subsection{Other Modules}

MXNet ships with tools to pack arbitrary sized examples into a single compact file to facilitate both sequential and random seek. Data iterators are also provided. Data pre-fetching and pre-processing are multi-threaded, reducing overheads due to possible remote file store reads and/or image decoding and transformation.

The training module implements the commonly used optimization algorithms, such as stochastic gradient descent. It trains a model on a given symbolic module and data iterators, optionally distributedly if an additional \texttt{KVStore} is provided.


\section{Implementation}
\label{sec:implementation}

\subsection{Computation Graph}

\begin{wrapfigure}[4]{r}[0pt]{.35\textwidth}
  \vspace{-70px}
  \includegraphics[width=\linewidth]{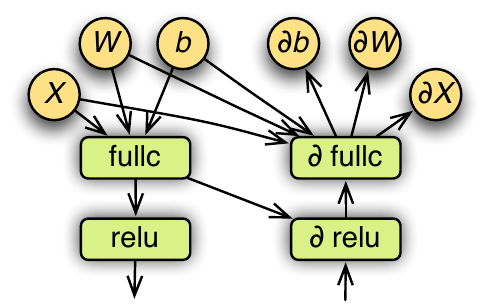}
  \vspace{-20pt}
  \caption{Computation graph for both forward and backward.}
  \label{fig:graph}
\end{wrapfigure}
A binded symbolic expression is presented as a computation graph for
evaluation. Figure~\ref{fig:graph} shows a part of the graph of both forward and
backward of the MLP symbol in Figure~\ref{fig:julia}. Before evaluation, MXNet
transforms the graph to optimize the efficiency and allocate memory to internal
variables.


\textbf{Graph Optimization.}
We explore the following straightforward optimizations. We note first that only the subgraph required to obtain the outputs specified during
binding is needed.
For example, in prediction only the forward graph is needed, while for extracting features from
internal layers, the last layers can be skipped.
Secondly, operators can be grouped into a single one. For example,
$a\times b+1$ is replaced by a single BLAS or GPU call.
%
Finally, we manually implemented well-optimized ``big'' operations, such as a
layer in neural network.

\textbf{Memory Allocation.}
%
Note that each variable's life time, namely the period between the
creation and the last time will be used, is known for a computation graph.
So we can reuse memory for non-intersected variables.
However, an ideal allocation strategy requires $O(n^2)$ time
complexity, where $n$ is the number of variables.

We proposed two heuristics strategies with linear time complexity.
The first, called {\em inplace}, simulates the procedure of traversing
the graph, and keeps a reference counter of depended nodes that are not used so
far. If the counter reaches zero, the memory is recycled.
%
The second, named {\em co-share}, allows two nodes to share a piece of memory if only if
they cannot be run in parallel. Exploring co-share imposes one additional
dependency constraint.
In particular, each time upon scheduling, among the pending paths in the graph, we find the longest path and perform needed memory allocations.

\subsection{Dependency Engine}

\label{sec:engine}

In MXNet, each source units, including \texttt{NDArray}, random number generator and temporal space, is registered to
the engine with a unique tag. Any operations, such as a matrix operation or data communication,
is then pushed into the engine with specifying the required resource tags.
The engine continuously schedules the pushed operations for execution if
dependencies are resolved.
%
Since there usually exists multiple computation resources such as CPUs, GPUs,
and the memory/PCIe buses, the engine uses multiple threads to scheduling the
operations for better resource utilization and parallelization.

Different to most dataflow engines~\cite{WanXiaLiZhaetal14}, our engine tracks mutation operations as an existing resource unit.
That is, ours supports the specification of the tags that a operation will \emph{write} in addition to \emph{read}.
This enables scheduling of array mutations as in numpy and other tensor libraries.
It also enables easier memory reuse of parameters, by representing parameter updates as mutating the parameter arrays.
It also makes scheduling of some special operations easier.
For example, when generating two random numbers with the same random seed, we
can inform the engine they will write the seed so that they should not be
executed in parallel. This helps reproducibility.

\subsection{Data Communication}

\begin{wrapfigure}[9]{r}{.3\columnwidth}
  \centering
  \vspace{-40pt}
  \includegraphics[width=\linewidth]{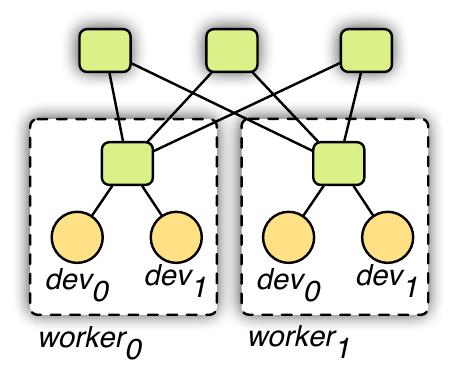}
  \vspace{-15pt}
  \caption{Communication. }
  \label{fig:ps}
\end{wrapfigure}

We implemented \texttt{KVStore} based on the parameter
server~\cite{LiAndParSmoetal14, LiAndSmoYu14, DeaCorMonCheetal12}(Figure~\ref{fig:ps}). It differs to previous works in two aspects:
First, we use the engine to schedule the \texttt{KVStore} operations and
manage the data consistency. The strategy not only makes the data synchronization
works seamless with computation, and also greatly simplifies
the implementation. Second, we adopt an two-level
structure.  A level-1 server manages the data synchronization between the
devices within a single machine, while a level-2 server manages inter-machine
synchronization. Outbound data from a level-1 server can be aggregated, reducing bandwidth requirement; intra- and inter-machine synchronization can use different consistency model (e.g. intra- is sequential and inter- is eventual).





\section{Evaluation}

\begin{figure}[t!]
  \centering
  \centering
  \begin{minipage}{.325\textwidth}
    \centering
    \includegraphics[width=.98\linewidth]{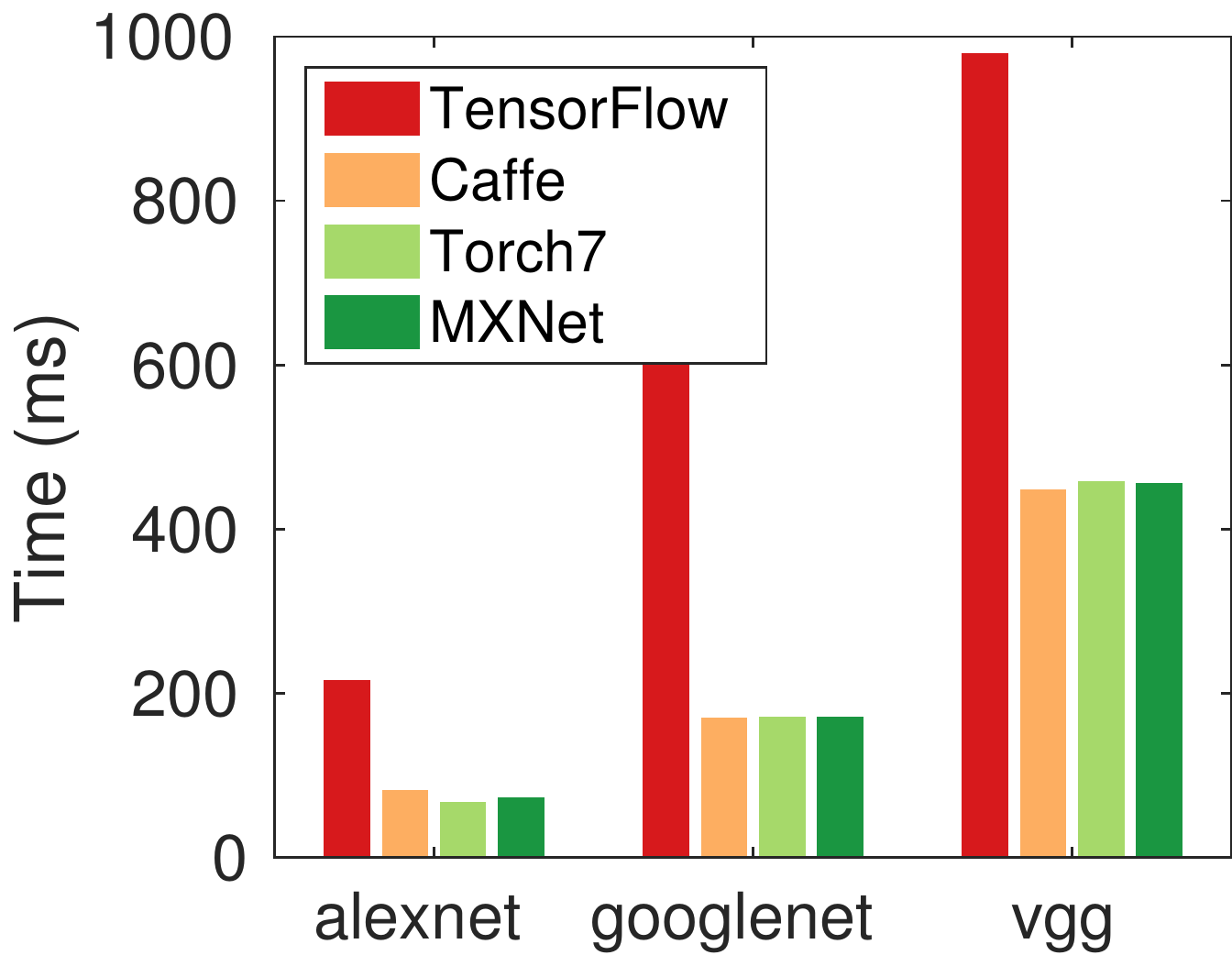}
    \caption{Compare MXNet to others on a single forward-backward performance.}
    \label{fig:conv_time}
  \end{minipage}\hfill%
  \begin{minipage}{.65\textwidth}
    \centering
    \includegraphics[width=.48\linewidth]{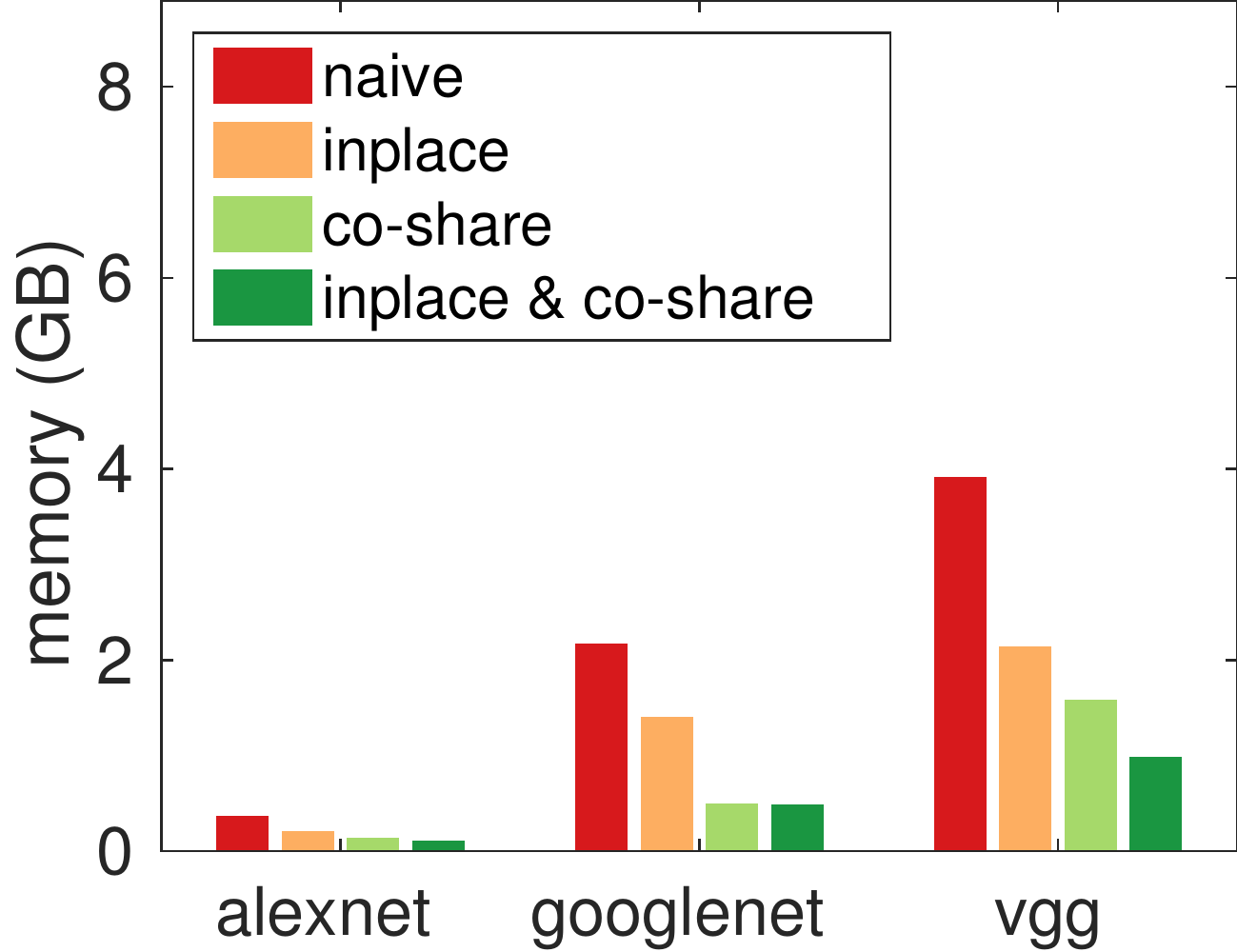}
    \includegraphics[width=.48\linewidth]{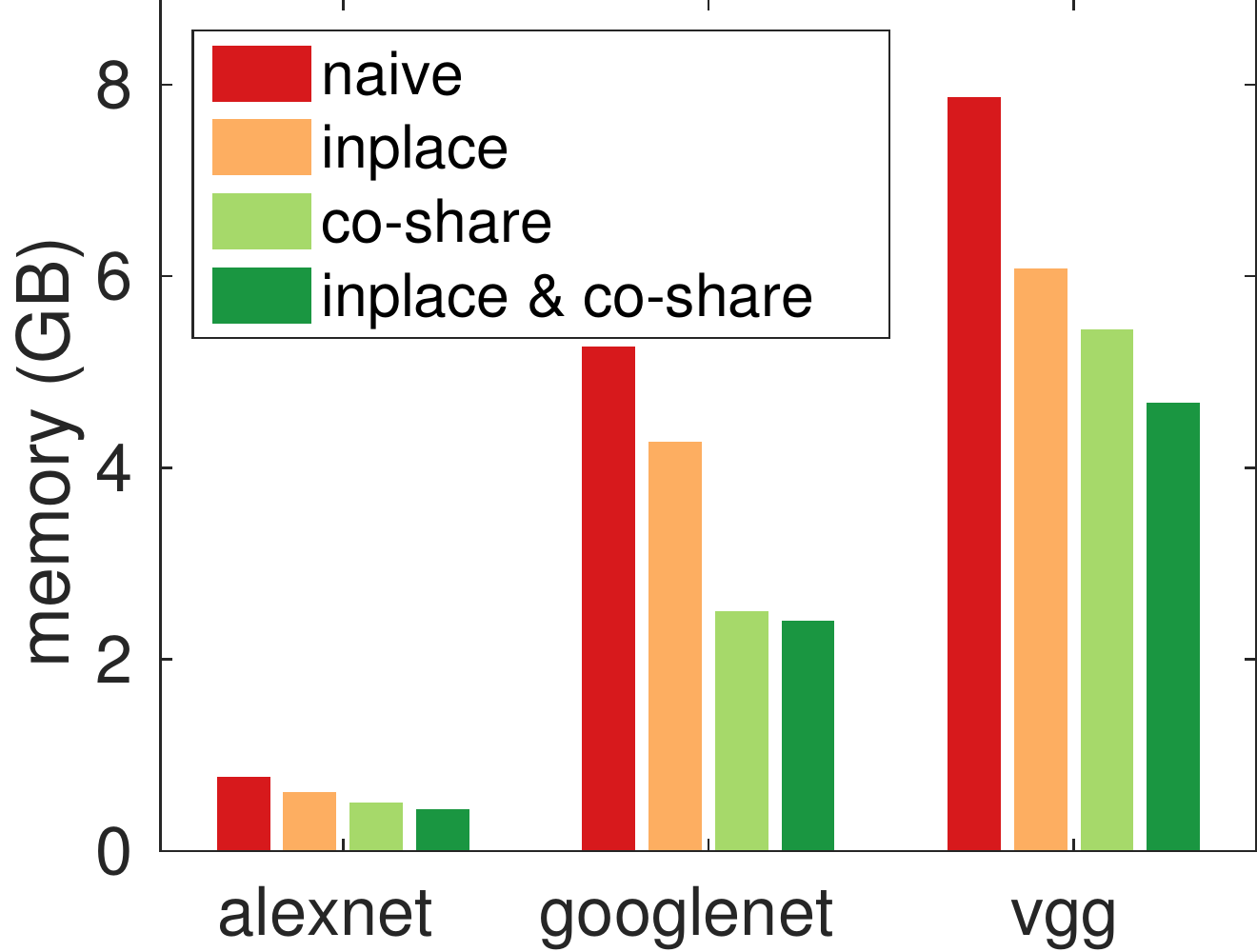}
    \caption{Internal memory usage of MXNet under various allocation strategies
      for only forward (left) and forward-backward (right) with batch size 64.}
    \label{fig:conv_mem}
  \end{minipage}
  \vspace{-4ex}
\end{figure}

\paragraph{Raw performance}
%
We fist compare MXNet with Torch7, Caffe, and TensorFlow on the popular
``convnet-benchmarks''~\cite{Chi15}. All these systems are compiled with CUDA 7.5 and
CUDNN 3 except for TensorFlow, which only supports CUDA 7.0 and CUDNN 2. We use
batch size 32 for all networks and run the experiments on a single Nvidia GTX
980 card. Results are shown in Figure~\ref{fig:conv_time}. As expected that
MXNet has similar performance comparing to Torch7 and Caffe, because most
computations are spent on the CUDA/CUDNN kernels.  TensorFlow is always 2x
slower, which might be due its use of a lower  CUDNN version.

\paragraph{Memory usage}
%
Figure~\ref{fig:conv_mem} shows the memory usages of the internal variables excepts for the outputs.
As can be seen, both ``inplace'' and ``co-share'' can effective reduce the memory
footprint. Combing them leads to a 2x reduction for all networks during model
training, and further improves to 4x for model prediction. For instance,
even for the most expensive VGG net, training needs less than {\em 16MB} extra.

\begin{wrapfigure}[15]{r}{6cm}
  \centering
  \vspace{-10px}
  \includegraphics[width=\linewidth]{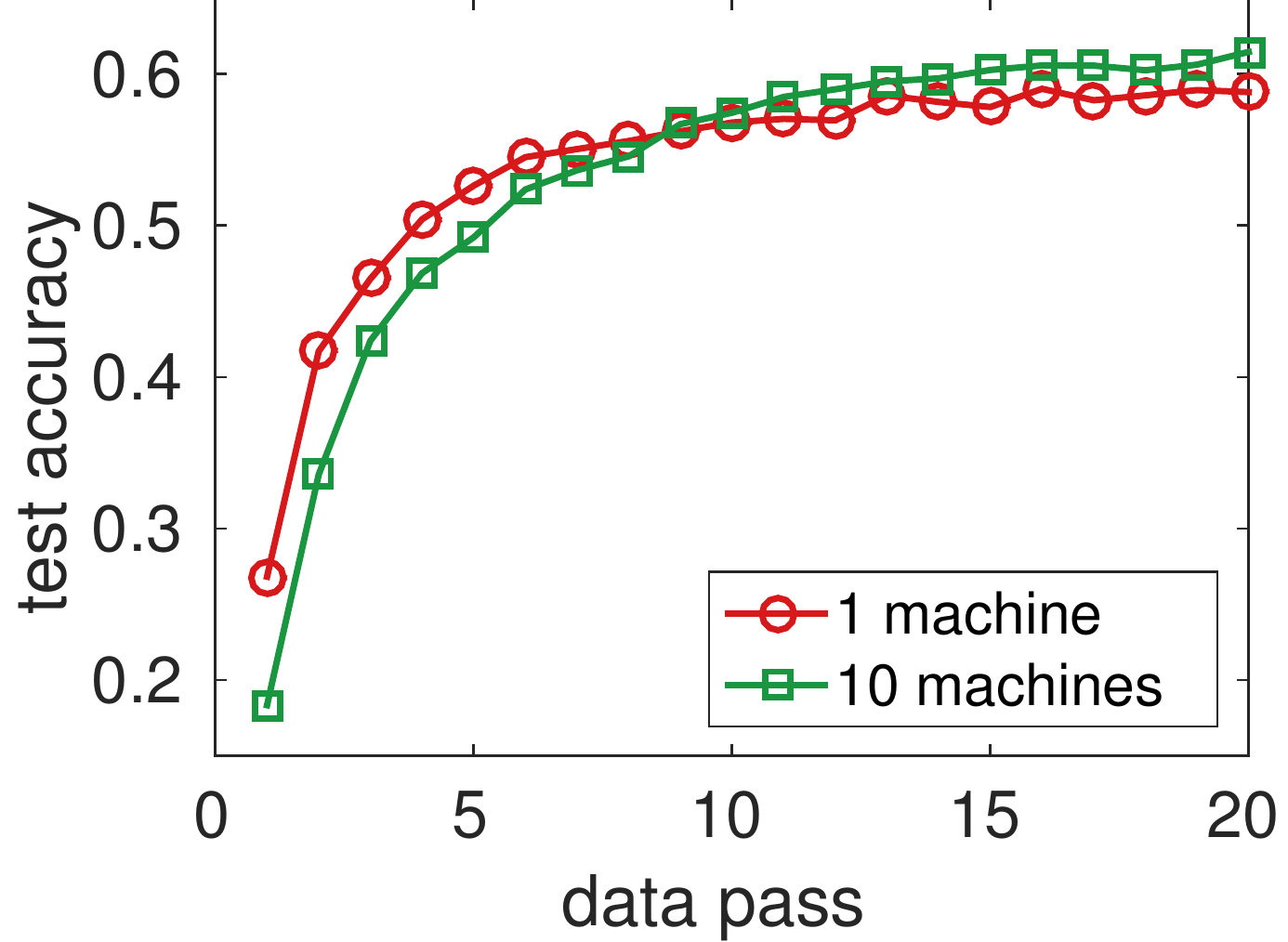}
  \vspace{-15px}
  \caption{Progress of googlenet on ILSVRC12 dataset on 1 and 10 machines.}
  \label{fig:dist_converge}
\end{wrapfigure}

\paragraph{Scalability}
%
We run the
experiment on Amazon EC2 g2.8x instances, each of which is shipped with four
Nvidia GK104 GPUs and 10G Ethernet. We train googlenet with batch
normalization~\cite{IofSze15} on the ILSVRC12
dataset~\cite{RusDenSuKraetal14} which consists of 1.3 million images and
1,000 classes. We fix the learning rate to $.05$, momentum to $.9$,
weight decay to $10^{-4}$, and feed each GPU with $36$ images in one batch.

The convergence results are shown in Figure~\ref{fig:dist_converge}. As can be seen, comparing to single machine,
the distributed training converges slower at the
beginning, but outperforms after 10 data
passes. The average cost of a data pass is 14K and 1.4K sec on a single machine and 10 machines, respectively. Consequently, this experiment reveals a super-linear speedup.

\vspace{-1ex}
\section{Conclusion}
\vspace{-1ex}
MXNet is a machine learning library combining symbolic
expression with tensor computation to maximize efficiency and
flexibility. It is lightweight and embeds in multiple host languages, and can be run in a distributed setting.
Experimental results are encouraging. While we continue to explore new design choices, we believe it can already benefit the relevant research community.
The codes are available at \url{http://dmlc.io}.

\textbf{Acknowledgment.}
We sincerely thanks Dave Andersen, Carlos Guestrin, Tong He, Chuntao Hong, Qiang
Kou, Hu Shiwen, Alex Smola, Junyuan Xie, Dale Schuurmans and all other contributors.

\bibliographystyle{plain}

\end{document}